\documentstyle[twoside,fleqn,espcrc2]{article}

\newcommand{\AmS}{{\protect\the\textfont2
  A\kern-.1667em\lower.5ex\hbox{M}\kern-.125emS}}

\title{The role of diffeomorphisms in the integration over a finite dimensional
space of geometries}

\author{Pietro Menotti \address{Dipartimento di Fisica della
Universit\`a di Pisa and INFN Sezione di Pisa\\
        Piazza Torricelli 2, 56100 Pisa, Italy}%
        \thanks{Work supported in part by MURST}
        }
\begin{document}

\begin{abstract}
Starting from the De Witt
supermetric and limiting ourselves to a family of geometries
characterized by
a finite number of geometric invariants we extract the unique
integration measure. Such a measure turns out to be a
geometric invariant, i.e.\ independent of the gauge fixed metric used
for describing the geometries. The measure is also invariant in form
under an arbitrary change of parameters describing the geometries.
The additional functional integration on the conformal factor makes
the measure
independent of the free parameter intervening in the De Witt
supermetric. The differences between the case $D=2$ and $D>2$ are
evidenced.
\end{abstract}

\maketitle

\section{INTRODUCTION}

We deal here with the regularization of the functional integral over
geometries, obtained by considering a finite dimensional space of
geometries i.e. by limiting the integration to a family of
geometries described by a finite number $N$ of geometric invariants which
we shall denote by $l_i$.

To specify the measure we must state
which is the basic integration variable. Following the lessons of
gauge theories we shall assume as basic integration variable the
metric $g_{\mu\nu}$ which shall play the role analogous to the field
$A_\mu$ of gauge theories. As in Wilson formulation
gauge invariance is treated exactly here the invariance under
diffeomorphisms will be dealt with exactly. Obviously this makes the
original degrees of freedom infinite and our purpose will be that the
derive a formula in which only the finite geometrical degrees of
freedom appear.

Essential is to state a diffeomorphism invariant distance among two
nearby field configurations and our choice will be the De Witt
supermetric \cite{dewitt}
$$
(\delta g,\delta g) = ~~~~~~~~~~~~~~~~~~~~~~~~~~~~~~~~~~~~~~~~~~~~~~~~~~~~~~~
$$
\begin{equation}
\label{dewittm}
\int \sqrt{g(x)}\, d^D x \,\delta
g_{\mu\nu}(x)
G^{\mu\nu\mu'\nu'}(x) \delta g_{\mu'\nu'}(x)
\end{equation}
where
$$
G^{\mu\nu\mu'\nu'}=~~~~~~~~~~~~~~~~~~~~~~~~~~~~~~~~~~~~~~~~~~~~~~~~~~~
$$
\begin{equation}
g^{\mu\mu'}g^{\nu\nu'} + g^{\mu\nu'}g^{\nu\mu'}-
{2\over D}
g^{\mu\nu}g^{\mu'\nu'}+C g^{\mu\nu}g^{\mu'\nu'}~~~~~~~~~~
\end{equation}
whose very well known property is to be the unique ultralocal
diffeomorphism invariant distance.

We recall that the De Witt distance in invariant only under ``rigid''
diffeomorphisms i.e. diffeomorphisms which are independent of the
parameters whose variations generate the $\delta g_{\mu\nu}$. We shall
see however that the final result will be invariant under the general
diffeomorphism.

A distance in the space of metrics induces a volume in the space of metrics
as it happens in finite dimensional spaces and this will be the origin of
the integration measure.

In the developments of sect.2  we shall need a positive definite
metric; this restricts our treatment to the Euclidean gravity with the
parameter $C$ appearing in the De Witt supermetric subject to the
restriction $C>0$.

Related ideas are found in \cite{jev} even if there are some
differences with respect to \cite{jev} and in \cite{mazm}  where
however one is concerned with the continuum problem.
Details of the material presented are found in \cite{pmppp}.

\section{INTEGRATION OVER THE DIFFEOMORPHISMS}

Everything that follows here will be a direct consequence of the De
Witt supermetric and of the restriction on the geometries to belong to a
finite dimensional family. We shall denote with $\bar
g_{\mu\nu}(x,l)$ a gauge fixed metric, i.e. a metric which describes
the geometry characterized by the invariants $l_i$; the final results
will be independent of
the choice of $\bar g_{\mu\nu}(x,l)$.
The general metric describing our geometries is given by
\begin{eqnarray}
g_{\mu\nu}(x,l,f) = [f^{\star}\bar g_{\mu\nu}(l)](x)\equiv \nonumber \\
\bar g_{\mu'\nu'}(x'(x),l)
{\partial x^{\mu'}\over \partial x^\mu}{\partial x^{\nu'}\over
\partial x^\nu}.
\end{eqnarray}
The problem is now to factorize in ${\cal D}[g_{\mu\nu}]$ the infinite
gauge volume and a term $J(l) \prod_{i=1}^N dl_i$.
Straightforward generalization of the finite dimensional procedure gives
for the Jacobian $J(l)$
\begin{equation}
\label{tang}
1=
J(l) \int\prod_i d\delta l_i {\cal D}[\xi]
e^{-{1\over 2} (\delta g,\delta g)}.
\end{equation}
In order to compute such a Jacobian it is useful to
decompose the general variation of the metric in two pieces which are
orthogonal according to the De Witt supermetric
\begin{eqnarray}
\delta g_{\mu\nu}(x) = F(\xi)_{\mu\nu} + \left[f^{\star} {\partial
\bar g_{\mu\nu}(l) \over dl_i} \delta l_i \right](x) \nonumber \\
= [ (F\xi)_{\mu\nu} + F(F^\dagger F)^{-1}
F^\dagger {\partial g_{\mu\nu}\over \partial l_i} \delta l_i ] +
\nonumber \\
\left[1 - F (F^{\dagger} F)^{-1} F^{\dagger}\right]
{\partial g_{\mu\nu}\over \partial l_i} \delta l_i
\end{eqnarray}
where
\begin{equation}
F(\xi)_{\mu\nu} = \nabla_\mu\xi_\nu + \nabla_\mu\xi_\nu.
\end{equation}
Substituting into (\ref{tang}) gives for the Jacobian $J(l)$
\begin{eqnarray}
\label{jacob}
J(l) =  \det (t^i, t^j)^{\frac{1}{2}} {\cal D}\mbox{et}
(F^\dagger F)^{\frac{1}{2}}
\end{eqnarray}
where
\begin{equation}
\label{proj}
t^i_{\mu\nu}=[1 - F (F^\dagger F)^{-1} F^\dagger]
{\partial g_{\mu\nu}\over \partial l_i}.
\end{equation}
Thus the measure (\ref{jacob}) is the product of two factors. The first
factor counts
the number of different \underbar{geometries} which belong to the element
$\prod_i dl_i$ while the second one is a true functional determinant giving
the gauge volume of each geometry, i.e. the number of
different \underbar{metrics}
which describe a fixed geometry characterized by the parameters
$l_i$. The second factor
is the analogous of the Liouville action in the treatment of two
dimensional gravity in the conformal gauge \cite{allc,pmppp}.

Both terms are invariant under diffemorphisms
also when we let them depend on the parameters $l_i$. In the proof of
such a property \cite{pmppp} the projector appearing on the l.h.s. of
eq.(\ref{proj}) plays an
essential role; as a result both terms are geometric invariants, i.e.
they are functions only of the $l_i$ and do not depend on the
original gauge fixed metric $\bar g_{\mu\nu}$.

Obviously due to (\ref{proj}) the expression $J(l)\prod dl_i$ is
invariant under a change of the set of parameters $l_i$ into any other
set of parameters which describe the same geometries. The source of
such invariance is that the metric $g_{\mu\nu}$ and not the parameters
$l_i$ have been chosen as fundamental integration variables.

As the $l_i$ are geometric invariants no diffeomorphism can connect
two points on the gauge fixing surface $\bar g_{\mu\nu}(x,l)$ and thus
no Gribov problem arises in this scheme.

In the Regge case if we give the $l_i$ the meaning of the link lengths,
it is well known \cite{hartle} that for a zero measure set of
values of the $l_i$, there are changes $\bar{\delta l_i}$ which leave the
geometry unchanged. Not only such a set of values is of zero measure
but in addition on such set
the term $\det(t^i,t^j)$ vanishes because in such a case a $\bar\xi$
exists such that
$
\frac{\partial \bar g_{\mu\nu}}{\partial_{l_i}}\bar\delta l_i = (\bar
F\bar\xi)_{\mu\nu}
$
which gives $t_i\bar{\delta l_i} =0$ and thus $\det(t^i,t^j)=0$.

Great simplifications would occur if a gauge could be found in which
\begin{equation}
\label{transverse}
\bar F^\dagger  \frac{\partial \bar g_{\mu\nu}}{\partial l_i}=0.
\end{equation}
The analogous problem in gauge theories
is to find a surface in the space of field configurations which is
orthogonal to all gauge fibers. Such a
surface in general, i.e. for a
sufficiently rich choice of $A(x,l)$ does not exist for non abelian
theories. This is also true in gravity where one can give in $D>2$
simple non
pathological examples of families of geometries for which
such a gauge fixing surface in the space of field configurations
does not exist (see
ref.\cite{pmppp} Appendix A). A gauge satisfying (\ref{transverse})
can be found only for simple
minisuperspace models.

As $F^\dagger$ depends on $C$ through the De Witt metric, both terms
in (\ref{jacob}) are $C$ dependent and one can show that  such a
dependence does not
cancel out \cite{pmppp}. As already shown in \cite{allc} such a $C$
dependence disappears
once one integrates over all conformal variations of our family of
metrics. Thus in addition to a finite number of parameters
$\tau_{i}$ which describe deformations transverse (i.e.\ non
collinear) to the orbits generated by the Weyl and
diffeomorphism groups, we consider the deformations induced by the
Weyl group \cite{mazm}.
Inclusion of the conformal factor produces the following changes:
the operator $F$ is replaced by its conformal version
$P$ defined by
\begin{equation}
P(\xi)_{\mu\nu}= F(\xi)_{\mu\nu} - \frac{g_{\mu\nu}}{D}
g^{\lambda\rho}F(\xi)_{\lambda\rho}
\end{equation}
and $\frac{\partial g_{\mu\nu}}{\partial \tau_i}$ is replaced by
\begin{equation}
k^i_{\mu\nu} =  \frac{\partial g_{\mu\nu}}{\partial \tau_{i}} -
\frac{g_{\mu\nu}}{D} g^{\alpha\beta} \frac{\partial g_{\alpha
      \beta}}{\partial \tau_{i}}
\end{equation}
reaching for the integration measure $J(\sigma, \tau)$
\begin{equation}
\label{sigmameasure}
\displaystyle
{\cal D}\mbox{et} ({P}^{\dag} {P}
  )^{\frac{1}{2}}
    \left[ \det \left({k}^{i}, ( 1 - {P}
   \frac{1}{{P}^{\dag}{P}}{P}^{\dag})
   {k}^{j} \right) \right]^{\frac{1}{2}}.
\end{equation}
The next job is to extract the dependence on the conformal factor
$\sigma$ of the two terms appearing in eq.(\ref{sigmameasure}),
induced by the
dependence of $P$
\begin{equation}
\label{sigmadep}
P = e^{2\sigma} \hat P e^{-2\sigma},
P^\dagger=e^{-D\sigma} \hat P^\dagger e^{(D-2)\sigma}
\end{equation}
being $\hat P$ the operator on the background metric. Here the first
fundamental
difference appears
between the case $D=2$ \cite{allc,pmppp} and $D>2$. $\mbox{Ker}(
P^{\dagger})$ is the
analogous of the pure Teichm\"uller deformations in two dimensions and
in $D>2$ the dimensions of $\mbox{Ker}(P^{\dagger})$ is infinite.
By using eq.(\ref{sigmadep}) the dependence of the second factor in
(\ref{sigmameasure})  on $\sigma$ can be given in terms
on an $N\times N$ matrix which however involves the properties of the
whole space $\mbox{Ker}(P^{\dagger})$. In $D=2$ instead
the dimension of such a subspace is always finite dimensional and the
dependence of such a factor on $\sigma$ can be taken into account
explicitly.

With regard to the term ${\cal D}\mbox{et} ({P}^{\dag}
{P})$ it is given by the usual $Z$-function expression
\begin{eqnarray}
\label{Det}
\log {\cal D}\mbox{et} ({P}^{\dag}{P}) = - \frac{d}{ds} Z(0)
=\nonumber \\
- \frac{d}{ds} \left[ \frac{1}{\Gamma (s)} \int_{0}^{\infty} \! dt \:
t^{s -1} \mbox{Tr} (e^{-t {P}^{\dag}{P}}) \right].
\end{eqnarray}
Being ${P}^{\dag}{P}$ an elliptic operator for any $D$,
expression (\ref{Det}) is mathematically well defined \cite{nash}.
The procedure which works in $D=2$ is to compute its variation with
respect to $\sigma$ and then to integrate back the result
\cite{allc,pmppp}. The
variation of eq.(\ref{Det}) is
\begin{eqnarray}
-\delta \log{\cal D}\mbox{et} (P^\dagger P)
= \gamma_E \delta Z_{ P^\dagger P}(0) +
{\mbox{Finite}}_{\epsilon \rightarrow 0} \nonumber \\
\{(2+D)\mbox{Tr}(e^{-\epsilon P^\dagger P}
\delta\sigma) -D  \mbox{Tr}'(e^{-\epsilon P P^\dagger}\delta\sigma)\}.
\end{eqnarray}
Here the second fundamental difference occurs: it is easily seen that
in $D=2$ not only the operator $P^\dagger P$ is elliptic but
also the operator $P P^\dagger$ appearing in the second term
is elliptic, while in $D>2$ $P P^\dagger$ is no longer
elliptic which makes the usual heat kernel technology inapplicable. Thus
while ${\cal D}\mbox{et} (P^\dagger P)$ is a mathematically
well defined object its explicit dependence on $\sigma$, for $D>2$ is
not jet known.

\end{document}